# Foundations of e-Democracy[1]

**Ehud Shapiro**
Department of Computer Science & Applied Math
Weizmann Institute of Science
Rehovot, Israel
ehud.shapiro@weizmann.ac.il

**ABSTRACT**

The Internet revolution of democracy, which will transform earthly representative democracies by employing the communication and collaboration capabilities of the Internet, has yet to come. Here, we enlist the wisdom of our forefathers to lead the way. By consulting the *1789 Declaration of the Rights of Man and of the Citizen*, we distill core values of democracy and derive from them requirements for the foundations of e-democracy. Building these foundations can usher the urgently-needed revolution of democracy.

## 1 INTRODUCTION

Representative democracy is in retreat world-wide[1-3], as many democracies transform into oligarchies, plutocracies or even kleptocracies. A key reason is lack of respect to democracy's basic tenet – equality of rights – as the rich, the powerful and the connected increasingly dominate who gets nominated, who gets elected and what the elected do. The forefathers of democracy have identified this to be "...the sole cause of public calamities and of the corruption of governments."[4]

The Internet, on the other hand, is revolutionizing industry after industry, leaving older ways of human conduct in the dustbin of history. Yet, it has not changed the basic workings of democracy: Representative democracy today functions essentially as it did two hundred years ago (Internet-enabled disruptions of elections notwithstanding).

How could this be? How come an Internet revolution of democracy has not happened yet, despite the pressing need for it and the apparent clear ability of the Internet to deliver it? I believe a key reason is that amalgamating "Internet" and "Democracy" into an Internet democracy, or *e-democracy*, is harder than it seems.

E-democracy has at least two meanings: (i) Using the Internet to strengthen real-world democracies[1, 5] and (ii) Democratic conduct of virtual Internet communities[6]. When viewed as objectives they coalesce, as one entails or requires the other.

Amalgamating "Internet" and "Democracy" presupposes universal Internet access as well as net neutrality and freedom; their absence undermines the legitimacy of e-democracy, as a regime can exclude an oppressed minority, or a service provider can make e-democracy a super-premium service, excluding the poor.

Even if the Internet infrastructure is universally accessible, neutral and fair, utilizing an existing Internet application such as Facebook and its siblings as a foundation for e-democracy is a non-starter: They are prone to duplicate and fake accounts and, crucially, to nondemocratic oversight, control and arbitrary intervention by their owners. Even Wikipedia, a hallmark of Internet participation, is governed neither by its readers nor by its editors, but by an appointed board that has full legal authority to shut it down, e.g. to avert bankruptcy.

Hence, new *foundations for e-democracy* are needed. We envision these foundations to simultaneously support the democratic conduct of all types of communities: Associations, clubs, unions, cooperatives, organizations, movements, and political parties; and at all levels – local, national, transnational, and international; eventually including cities, states and federations; and, ultimately, uniting the entire humanity in a global e-democracy.

Among these communities, the pivot for revolutionizing earthly democracies may be Internet-resident democratic political parties, or e-parties. Only by winning real-world elections, e-parties can export the participatory practices of e-democracy from their inner workings to real-world governments, enacting

---

legislation that gradually supplants traditional representative democracy by e-democracy.

But what are these foundations?  Who could guide us in their construction? A standard method in requirements engineering is to interview the prospective customer. The prospective "customer" for e-democracy is humanity at large. Hence,  in lieu of an interview, we enlist one of humanity's most inspiring documents: The 1789 French *Declaration of the Rights of Man and of the Citizen[4]* (Henceforth: Declaration),  which offers a concise, clear and bold expression of the essence of democracy. We study its Articles, extract from them core democratic values, and derive from these values requirements for the foundations of e-democracy.

## 2   CORE VALUES OF DEMOCRACY

We list core democratic values extracted from the Articles (marked by **A**) of the Declaration (Interpreting *Man*→ *Person*, *Citizen*→*Member*, and *Nation*→*Community*):

**1. Sovereignty**: The Declaration's **Article III (A3)** states that "The principle of any sovereignty resides essentially in the Nation. No body, no individual can exert authority which does not emanate expressly from it."  We interpret this principle to mean that **the members of an e-democracy are its sovereign.**

**2. Equality**: **A1** states that "Men are born and remain free and equal in rights. …".  Together with **A3** they imply that sovereignly must be equally shared, often stated as *one person – one vote*. But there is more to equality than the right to vote. **A4** states that the law is the expression of the general will and that all people have the right to contribute to its formation; and equally so, according to **A1**.  **A6** further states that all people, being equal in the eyes of the law, are equally admissible to all public posts. Equality extends not only to rights but also to obligations: **A12–14** ascertain the need for public services and for equally sharing their financing among members, but progressively, according to their ability to pay.

To summarize, all members of a democracy must have equal capacity to act as voters, discussants, proposers and public delegates, as well as share progressively the burden of public expenditures.

**3. Freedom**:  **A1** states that "men are born and remain free".  The nature of this freedom is further elaborated in other articles: **A10-11** espouse the **freedom of expression** within the limits if the law.  **A5** proclaims the freedom to take any action that is not harmful to others. Among those implied freedoms we note the **freedom of assembly[7]**, granting any group of people the freedom to assemble, and the **subsidiary principle**, granting such a group the freedom to make decisions that pertain to them.

**4. Transparency**: **A14-15** require that the conduct of public agents and the collection and expenditure of public funds be transparent.  Furthermore, **A2** states that the goal of any political association must be the conservation of the natural and imprescriptible rights of man: **liberty, property, safety and resistance against oppression**.  This can be ascertained in an ecosystem of e-democracies only if the decisions of each are transparent to the others.

**5**. **Property and Privacy**:  **A17** recognizes the right for property and its private use, which, extended to our times, incorporates the right for the ownership and privacy of information. The right to safety and resistance against oppression (**A2**) entails voter privacy to resist coercion.

**6. Justice**: Revolt against unjust rulers was crucial to the emergence of democracy, and justice is the focus of early charters of democracy such as the English Magna Carta[8] and the French Declaration. Indeed, **A1** and **A4-13** address the equal and just conception, application and enforcement of the law.  Furthermore, **A16** states that a **constitution** is needed to guarantee the rights of citizens and the separation of the powers of government.

## 3   REQUIREMENTS OF FOUNDATIONS OF E-DEMOCRACY

We now aim to derive from these core democratic values requirements for the foundations of e-democracy.

**1. Sovereignty:** Internet communities today, from the local bulletin board to almighty Facebook, are dictatorial, with an omnipotent administrator who determines who gets in, who is thrown out and what actions may each member take.  The administrator also has the capacity to shut down the community and annihilate its recorded history at will. Furthermore, communities like Facebook employ rule-by-decree like bygone Middle Ages fiefdoms.  The owner, like a feudal lord, sets the rules (sometimes in secrecy), tries members for breaching them and executes the punishment.  The members, like serfs, toil for the financial benefit of the lord while having no (intellectual) property, civil rights, or voting rights. They have no say on their remuneration or tax, on community rules of conduct or their enforcement, or on the election of community leadership. In the event of a bankruptcy or hostile takeover, the entire community and its recorded history may be annihilated, with community members being helpless bystanders. All this clearly violates all of democracy's core values – sovereignty, equality, freedom, transparency, property, privacy and justice.

First, we consider the question of ownership. Any seemingly sovereign e-democracy that resides on computers operated by a third party could be unplugged at its will, or its default, rendering sovereignty meaningless. Hence, in the context of an e-democracy, sovereignty requires ownership.

How can the members of an e-democracy be the sovereign and hence necessarily the owner? Advances in cryptocurrencies and blockchain technology provide the first example. In a DAO (Decentralized Autonomous Organization)[6], built on top of Ethereum, the dictatorial system administrator is replaced by a *smart contract*, namely an autonomous, incorruptible, transparent and persistent software agent, programmed to obey democratic decisions (albeit with *one coin – one vote*, not *one person – one vote*). The DAO operates on a distributed computer network with no central ownership. A few caveats: First, an early DAO venture capital fund had a bug that allowed a malicious member to syphon its funds. Smart-contract programming in general and the DAO architecture in particular have yet to mature to offer a sound foundation for e-democracy. Second, Ethereum and Bitcoin, while having distributed control in theory, have a core group of miners that could control and subvert them should they decide to join forces, a risk that a future e-democracy at the national or global scale cannot afford. Third, current proof-of-work consensus protocols of public blockchains incentivize inconceivable and unsustainable waste of energy, which cannot be endorsed by any moral person or organization. Fourth, a replicated ledger such as Ethereum and Blockchain could not support the high-throughput transaction rate and response time required by a national or global e-democracy; a distributed ledger architecture is needed. Fifth, to foster participation rather than greed, a democratic cryptocurrency should reward participation[9], rather than capital-intensive coin-mining; the globally-unique digital identities required for e-democracy, discussed below, may afford egalitarian cryptocurrency mining[9, 10].

The economy of a democratic cryptocurrency could be programmed with democratically instituted taxes and budgets (e.g., [11, 12]) to operate the e-democracy.

In summary, a distributed public ledger employing a democratic cryptocurrency and programmed to adhere to democratic control could ensure that the members of an e-democracy are its sovereign and owner.

**2. Equality:** Equality entails *one person – one vote*. Yet e-democracies consist of digital identities, not people. Requiring *one digital identity – one vote* is not enough, as most existing systems allow a person to create as many digital identities as one wishes.

To support equality in an e-democracy, a new notion of digital identity must be devised that is truthful, unique, persistent, and owned by the person it represents. Otherwise, if fake – the owner may vote on behalf of a non-existent person; if non-unique – the owner may cast multiple votes; if not persistent – the owner may terminate and shed an obligated identity and acquire a fresh one clear of obligations, eluding accountability; and if not owned by the person it represents – it grants its owner an extra vote at the expense of the person it represents.

While truthfulness is a common requirement, e.g. in credit card and mobile phone contracts, uniqueness and persistency are not, as a person may obtain numerous credit cards, mobile phones and email accounts and terminate them at will. Government-issued identity numbers, often complemented with biometric attributes and incorporated in digital identity cards (cf. e-Estonia, India's Aadhaar) may serve as a unique and persistent digital identity attribute.

However, e-democracies may transcend national boundaries, e.g. in regional and international organizations. Realizing equality in global e-democracies is a bigger challenge: First, unhindered Internet access should be a recognized basic civil right and be provided universally. Second, some people, notably refugees, may have no verifiable national identity, yet should be granted participation in a global e-democracy. Third, people may have multiple citizenships, and without an additional notion of "global citizenship" with an associated globally-unique digital identity, one may have multiple votes, violating equality. Fourth, malicious nondemocratic regimes may produce an arbitrary number of fake national identities and use them (in a Sybil attack[13]) to sway the vote of a global e-democracy in favor of their national interest.

A trustworthy notion of *global citizenship*; a mechanism to endow each global citizen with a truthful, persistent and globally-unique *global digital identity*; and a *global judiciary* empowered to revoke fake or duplicate global digital identities and to transfer stolen identities back to their rightful owners, as well as to prosecute the perpetrators of these crimes, are all needed to ensure equality in a global e-democracy.

**3. Freedom:** As **freedom of expression** is granted within the limit of the law, its realization requires a *constitution* that determines these limits and a *judiciary* that enforces them, discussed below. **Freedom of assembly** can by realized by a software architecture that allows the unhindered formation of one e-democracy within another. To uphold the **subsidiary principle**, each subsidiary democracy should be able to undertake decisions that pertain to it, within the law.

**4. Transparency:** The structure of an e-democracy, its rules of conduct, its underlying technology, the source code of its software, as well as the decisions of its communities, the actions of its public delegates and its finances must all be transparent to all. (We acknowledge that in an extreme scenario, resisting an oppressive regime may necessitate compromised transparency.)

**5. Property & Privacy:** The ownership of private data and its measured disclosure to third parties only as needed can be supported with self-sovereign identities[14]. Ensuring privacy of voters and avoiding coercion require advanced cryptographic techniques such as anonymous credentials[15] and coercion-avoidance.

**6. Justice and Accountability:** To advance from the Internet Middle Ages and supplant Internet fiefdoms with e-democracies, we must offer **justice** – a democratic mechanism for establishing fair law and order. Its components must be a *constitution*, subject to democratic amendment, and a democratically-elected *judiciary* that rules according to the constitution.

e-democracies will come under criminal attack through identity forgery and theft, voter coercion, misinformation, hate crimes, and other offences. They can be redressed by the judiciary via a public warning, public condemnation, temporary gag, and fines. As suspension or, worse, expulsion, violate the basic civil right to vote, it may be considered too extreme. Imagine a future in which a person is a member of multiple e-democracies, which have a joint judicial system. A temporarily limit on participation in all these democracies simultaneously, analog to jail time in the real world, may be severe indeed. But for such a punishment to be effective, **accountability** must be ensured: it is not sufficient that the offending digital identity be truthful; it has to be unique and persistent, lest the offender sheds the punishment by abandoning one identify in favor of another.

**7. Hysteresis:** Democracy's forefathers have not foreseen the immediacy with which the general will can be ascertained on the Internet. Eventually, the general will must prevail lest we violate sovereignty. But it should go through reasonable checks and balances until it does, lest mob dynamics prevail. To this end we enlist **hysteresis**, a characteristic of systems in which the output is not an immediate function of the input.

While a multi-year election cycle confers natural hysteresis on earthly democracies, e-democracies require hysteresis to be engineered, so that swings in people's opinions may not immediately result in decisions that accommodate such swings. Example are minimal periods for proposal making and deliberation; minimal endorsements for proposals to be considered; minimal quorum for a decision to be binding; and special majority needed for certain actions, e.g. change of constitution.

## 4   CONCLUSIONS

It is my opinion that representative democracies are in dire straits because of their failure to uphold core democratic values, notably equality and transparency, and that e-democracy may offer the only feasible remedy. We have derived requirements for the foundations of e-democracy from the 1789 French Declaration of the Rights of Man and of the Citizen. The next urgent step is to build such foundations so that the desperately-needed Internet revolution of representative democracy would commence.

**ACKNOWLEDGMENTS**

Ehud Shapiro is the incumbent of the Harry Weinrebe Chair of Computer Science and Biology. Thanks to Ofir Raz, Amos On, Luca Cardelli, Jeffrey Sachs, Stan Letovsky, Yaniv Erlich, Benny Daon, Shani Gershi, Nimrod Talmon, Ariel Procaccia, Liav Orgad and Raffaele Marchetti for discussions and helpful comments and to Michal Golan-Mashiach and Ouri Poupko for their help.

**REFERENCES**

# Foundations of e-Democracy

[1] Anttiroiko, A.-V. Building strong e-democracy: the role of technology in developing democracy for the information age. *Commun. ACM*, 46, 9 (2003), 121-128.
[2] Democracy Index - Wikipedia. *https://en.wikipedia.org/wiki/Democracy_Index* (2017).
[3] Foa, R. S. and Mounk, Y. The Danger of Deconsolidation; The Democratic Disconnect, 27, 3 (2016).
[4] Déclaration des droits de l'homme et du citoyen; National Constituent Assembly. See *https://en.wikipedia.org/wiki/Declaration_of_the_Rights_of_Man_and_of_the_Citizen for an English translation.* (1789).
[5] Schuler, D. Creating the world citizen parliament: seven challenges for interaction designers. *Interactions*, 20, 3 (2013), 38-47
[6] *Create a Democracy contract in Ethereum*. https://www.ethereum.org/dao, 2017.
[7] Mill, J. S. *On Liberty*. Longman, Roberts & Green, Bartleby.com, 1999, 1859.
[8] Magna Carta Libertatum. *https://en.wikipedia.org/wiki/Magna_Carta* (1215).
[9] Jeffries, D. Why Everyone Missed the Most Mind-Blowing Feature of Cryptocurrency. *Hacker Noon* (2017).
[10] Gilad, Y., Hemo, R., Micali, S., Vlachos, G. and Zeldovich, N. Algorand: Scaling Byzantine Agreements for Cryptocurrencies. *Cryptology ePrint Archive*, Report 2017/454 (2017).
[11] Lee, D. T., Goel, A., Aitamurto, T. and Landemore, H. Crowdsourcing for Participatory Democracies: Efficient Elicitation of Social Choice Functions. In *Proceedings of the Second AAAI Conference on Human Computation and Crowdsourcing (HCOMP-14)* (Pittsburgh, Pennsylvania, 2014). AAAI Press, [insert City of Publication],[insert 2014 of Publication].
[12] Shapiro, E. and Talmon, N. A Condorcet-Optimal Participatory Budgeting Algorithm. *arXiv* (2017).
[13] Brian Neil, L., Clay, S. and Margolin, N. B. *A Survey of Solutions to the Sybil Attack*. University of Massachusetts Amherst, Amherst, MA, 2006.
[14] Lewis, A. A gentle introduction to self-sovereign identity (2017).
[15] Camenisch, J. and Lysyanskaya, A. An Efficient System for Non-transferable Anonymous Credentials with Optional Anonymity Revocation. *Advances in Cryptology — EUROCRYPT 2001*, 2045 (2001).